\documentclass[11pt]{article}
\usepackage{amsfonts,amsmath,amssymb}
\usepackage{graphicx}
\usepackage{epic,eepic,color,graphicx}

\newfont{\twelvemsb}{msbm10 scaled\magstep1}
\newfont{\eightmsb}{msbm8}
\newfam\msbfam
\textfont\msbfam=\twelvemsb \scriptfont\msbfam=\eightmsb
\catcode`\@=11
\def\Bbb{\ifmmode\let\next\Bbb@\else
\def\next{\errmessage{Use \string\Bbb\space only in math mode}}\fi\next}
\def\Bbb@#1{{\fam\msbfam{{#1}}}}

\newcommand{\be}{\begin{equation}}
\newcommand{\ee}{\end{equation}}
\newcommand{\ba}{\begin{eqnarray}}
\newcommand{\ea}{\end{eqnarray}}

\begin{document}
\sloppy
\renewcommand{\thefootnote}{\fnsymbol{footnote}}
\newpage
\setcounter{page}{1} \vspace{0.7cm}

\vspace*{1cm}
\begin{center}
{\bf On the origin of the correspondence between classical and quantum integrable theories}\\
\vspace{1.8cm} {\large Davide Fioravanti $^a$ and Marco Rossi $^b$
\footnote{E-mail: fioravanti at bo.infn.it, rossi at cs.infn.it}}\\
\vspace{.5cm} $^a${\em Sezione INFN di Bologna, Dipartimento di Fisica e Astronomia,
Universit\`a di Bologna\\
Via Irnerio 46, 40126 Bologna, Italy} \\
\vspace{.3cm} $^b${\em Dipartimento di Fisica dell'Universit\`a della Calabria and
INFN, Gruppo collegato di Cosenza\\
Arcavacata di Rende, 87036 Cosenza, Italy} \\
\end{center}
\renewcommand{\thefootnote}{\arabic{footnote}}
\setcounter{footnote}{0}
\begin{abstract}
If we start from certain functional relations as definition of a quantum integrable theory, then we can derive from them a linear integral equation. It can be extended, by introducing dynamical variables, to become an equation with the form of Marchenko's. Then, we derive from the latter a classical (differential) Lax pair. We exemplify our method by focusing on the massive version of the ODE/IM (Ordinary Differential Equations/Integrable Models) correspondence from Quantum sine-Gordon (sG) with many moduli/masses to the classical sinh-Gordon (shG) equation, so describing, in a particular case, some super-symmetric gauge theories and the $AdS_3$ strong coupling scattering amplitudes/Wilson loops. Yet, we present it in a way which reveals its generality of application. In fact, we give some hints on how it works for spin chains.
\end{abstract}
\vspace{1cm} 
\newpage

\section{Introduction}
\label{intro}
\setcounter{equation}{0}

One remarkable correspondence of modern mathematical physics is the so-called ODE/IM correspondence \cite{ODE-IM}. In a nutshell, it starts from the monodromies of a suitable Schr\"odinger equation and surprisingly derives the eigenvalues of two celebrated Baxter operators, the $Q$ and $T$ (functions), in the case of 2D Conformal Field Theories (CFTs). The natural evolution of this correspondence, with moduli, i.e. masses, and for the ground state, has been proposed by \cite{GMN} and then \cite{LZ}: they introduced first order differential $2\times2$ matrix operators $D$, $\bar D$, instead of second order differential scalar one, and studied monodromies of the solutions of the Lax linear problems
\be
D\Psi =0 \, ,\,\,\,  \bar D \Psi=0 \, , \label{Laxpr}
\ee
with $D$ and $\bar D$ given by
\be
D=\frac{\partial}{\partial w}+\frac{1}{2}\frac{\partial \hat \eta }{\partial w}\, \sigma ^3 -e^{\theta +\hat \eta  } \sigma ^+ - e^{\theta -\hat \eta  } \sigma ^- \, , \quad \bar D=\frac{\partial}{\partial \bar w}-\frac{1}{2}\frac{\partial \hat \eta  }{\partial {\bar w}}\, \sigma ^3 -e^{-\theta +\hat \eta  } \sigma ^- - e^{-\theta -\hat \eta  } \sigma ^+ \, , \label{conn}
\ee
with $\hat \eta (w,\bar w)$ a 2D classical scalar field. It satisfies the zero curvature condition $[D,\bar D]=0$, which happens to be the classical sinh-Gordon (shG) equation in this case. Crucially, the coefficients of these monodromies satisfy many functional relations \cite{LZ}: $QQ$-, $TQ$- and then $T$- and $Y$-systems, or equivalently the Thermodynamic Bethe Ansatz (TBA) equations of some integrable Quantum Field Theory (QFT), in this case sine-Gordon (sG) and generalisations with many mass scales \cite {FRS}. As for the last cases, they introduced in the Lax operators a set of parameters, the moduli ($\vec{c}$ below), which turn out, in the end, to parametrise the masses of the QFT. What was lacking in this scenario was a systematic way to understand how to derive, in {\it the opposite direction}, a classical system from a quantum one: in this letter we propose a procedure to realise this program. In other words, we want to start from a quantum integrable model and then derive in a precise way a classical model associated to it. In fact, we start from a precise definition of an integrable system (field or lattice theory) in terms of $Q$ functions (eigenvalues of $Q$ operators) and functional relations satisfied by them, the so-called $QQ$-system. Along the lines of \cite {FRS} all the other integrable structures, Baxter's and \textit{universal} $TQ$-system, TBA \cite {TBA} and Non Linear Integral Equations for the counting functions \cite {NLIE}, the $T$- and $Y$-systems, can be derived. Yet, the main idea below is that we need only to convert the universal $TQ$-system into a linear integral equation and then, by suitable Fourier transform, to a not always well-defined Volterra integral equation. At this point we shall naturally introduce the space $w,\bar w$ of the ODE/IM correspondence as dynamical extension of the adimensional product $r=MR$ of the (lightest) mass $M$ and the circumference length $R$ (of the cylinder)\footnote{The CFTs will be obtained as a scaling limit for the mass going to zero in section \ref{cft case}.}. In this way, the integral equation becomes well-defined and acquires the form of that of Marchenko (formally without the bound state sum)~\cite {M}, named below Marchenko-like equation. Eventually, from this equation we derive two Schr\"odinger equations or two classical linear Lax problems, indeed equivalent to (\ref  {Laxpr}, \ref {conn}) above. We provide the details for this specific case, which also describes, in a particular case, scattering amplitudes or null polygonal Wilson loops in $\mathcal{N} = 4$ SYM at strong coupling. In this regime they are indeed dominated by the string classical contribution, i.e. the minimal area of the
surface in $AdS_3$ ending on the boundary on the Wilson loop and they are essentially given by the free energies of the quantum model associated to (\ref  {Laxpr}, \ref {conn}) \cite{Ys}. But we can ostensibly adapt the method to general cases. To corroborate this statement, we give here only hints on spin chains by developing our construction for the $XXZ$ at $\Delta=-1/2$ and easily yielding Stroganov's equations \cite{STR}. 


\section{From quantum to classical theory exactly (without approximation)}
\label{Q-GLM}
\setcounter{equation}{0}

In the case of scattering amplitudes/Wilson loops in $\mathcal{N} = 4$ SYM at strong coupling our starting hypothesis is the $QQ$-system (2.20) of \cite{FRS}\footnote{In relation (2.20) of \cite{FRS} an extra phase factor, denoted $e^{i\Phi \left (\theta + \frac{i\pi}{2N} , \vec{c}\right )}$, with $\Phi$ different from zero only in particular cases,  is also present. Inclusion of this factor will not alter the conclusions of this letter, but would render notations more unwieldy. Therefore, for clarity's sake, we decided to omit it in this brief note.},
\be
Q_+ \left (\theta + \frac{i\pi}{2N}, \vec{c} \right ) Q_- \left (\theta - \frac{i\pi}{2N}, \vec{c}^R \right )- Q_+ \left (\theta - \frac{i\pi}{2N}, \vec{c}^R\right )Q_- \left (\theta + \frac{i\pi}{2N}, \vec{c} \right )=-2i  \cos \pi l  \label {Qw} \, ,
\ee
which connects two eigenvalues $Q_{\pm} (\theta ,\vec{c})$, the so-called $Q$-functions, of a $Q$ (Baxter) operator. We have that $|2l|<1$, $2N$ is a positive integer, $\theta $ is the so-called spectral parameter and $Q_{\pm} (\theta ,\vec{c})$ are, by assumption, entire functions of $\theta$, from which, in our perspective, all the physical quantities ensue. In fact, from (\ref {Qw}) we can derive Bethe equations, the $TQ$-, $T$- and the $Y$-systems as we have sketched in \cite{FRS} and will summarise below. In writing (\ref {Qw}) we made the hypothesis that $Q_{\pm}$ depend also on a vector of $2N-1$ complex parameters $\vec{c}=(c_0,...,c_{2N-2})$, the moduli, and on its 'rotated' version:
\be
\vec{c}\rightarrow \vec{c}^R =(c_0,..., c_n e^{-i \pi \frac{n}{N} }, ..., )
\label{crot} \, .
\ee
In the usual direct perspective of \cite {FRS} the transformation (\ref{crot}) is part of the so-called $\hat \Omega$-symmetry (see also the beginning of section \ref{cft case}). Another important assumption is the quasi-periodicity of the functions
\be
Q_{\pm } \left (\theta -i\tau , \vec{c} ^R \right )=e^{\mp i\pi \left (l+\frac{1}{2}\right )}  Q_{\pm} (\theta ,  \vec{c} ) \label {qper} \, ,
\ee
with period $\tau=\pi +\pi/N$. Its application removes from (\ref {Qw}) the appearance of the 'rotation', so that eventually we are left with a {\it universal} form for the $QQ$-system,
\be
e^{i\pi l} Q_+(\theta ,\vec{c})Q_-(\theta +i\pi, \vec{c}) + e^{-i\pi l}Q_-(\theta ,\vec{c} )Q_+(\theta +i\pi, \vec{c})=-2\cos \pi l
\label{QQ} \, ,
\ee
where also the shifts are fixed (and no longer depending on $N$). To continue the comparison with  \cite{FRS} in reverse order, this (via the above quasi-periodicity) has been obtained there by using the invariance of (\ref{Laxpr}) under the so-called $\hat \Pi$-symmetry.

\medskip

Now we move from (\ref {QQ}) and derive many functional and integral relations which eventually yield the associated Lax problem (\ref{Laxpr},\ref{conn}). First of all, we introduce a very useful quadratic construct of $Q_{\pm}$, the transfer matrix eigenvalue:
\be
T(\theta , \vec{c})=\frac{i}{2\cos \pi l} \left [ e^{-2i\pi l} Q_+(\theta +i\pi, \vec{c})Q_-(\theta -i\pi, \vec{c})-e^{2i\pi l} Q_+(\theta -i\pi, \vec{c})Q_-(\theta +i\pi, \vec{c}) \right] \label{Tdef} \, .
\ee
Combining relation (\ref {Qw}) with the quasi-periodicity (\ref {qper}) we arrive at the functional relation
\be
T(\theta, \vec{c})  Q_{\pm}(\theta, \vec{c} ) = Q_{\pm}\left(\theta +i\tau-i\pi, \vec{c}^{R^{-1}}\right)+Q_{\pm}\left(\theta -i\tau +i\pi, \vec{c}^{R}\right) \, ,
\label{TQ2}
\ee
with shifts depending on $\tau $. Relation (\ref{TQ2}) is the usual form of the Baxter $TQ$-relation for integrable models. However, for our aims it is more convenient to combine (\ref {QQ}) with (\ref {Tdef}) to arrive to the relation
\be
 T (\theta , \vec{c})  Q_{\pm}(\theta , \vec{c} )= e^{\mp i\pi \left (l+\frac{1}{2}\right) } Q_{\pm}(\theta +i\pi, \vec{c})+e^{\pm i\pi \left (l+\frac{1}{2}\right)} Q_{\pm}(\theta -i\pi, \vec{c}) \, , \label {TQstart}
\ee
which is a new $TQ$-system in a {\it universal} form, in the sense that the moduli do not rotate (and thus is more effective in their presence) and the shifts on the spectral parameter do not depend on $N$. Relation (\ref {TQstart}) expresses the transfer matrix $T$ via the Baxter auxiliary functions $Q_{\pm}$, or in reverse yields $Q_{\pm}$ as solutions of a finite difference second order equation, given the 'potential' $T$. And {\it en passant} we remark that (\ref {TQstart}, \ref {qper}) constrain it to be periodic
\be
T\left (\theta + i\tau, \vec{c} \right )=T(\theta , \vec{c}^{R})
\label{Tper} \, ,
\ee
so that the $Q_{\pm}$ are the Floquet solutions. In the end, it is also quite natural to assume the $Q$ and $T$ functions to be real-analytic (bar represents complex conjugation):
\be
\bar Q_{\pm} (\theta , \vec{c} )=
Q_{\pm} (\bar \theta , \bar \vec{c} )
\, , \quad \bar T(\theta , \vec{c}) =T (\bar \theta , \bar \vec{c})
\label{TQreal-an} \, .
\ee
The two properties (\ref {TQstart}, \ref {qper}) are general and to be specific they have to be equipped with the asymptotic
behaviour of $Q_{\pm}$, which usually fixes the integrable model and  also the state thereof. To make a definite example we choose the asymptotic behaviour typical for {\it the ground state} of massive integrable quantum field theories (though we may introduce that for an excited state and carry on similarly):
\be
\lim _{\textrm{Re} \, \theta \rightarrow \sigma \infty} \ln \left [ Q_{\pm}\left (\theta +i\frac{\tau}{2}, \vec{c}\right ) \right ] =  -w_0^{(\sigma)} (\vec{c}) e^{\sigma \theta}+ O(1)
\label{Qasy} \, , \quad |\textrm{Im} \theta |<\frac{\tau}{2} \, ,
\ee
where the (adimensional) {\it renormalisation group (RG) times} $w_0^{(1)}(\vec{c})=w_0(\vec{c})$ and $w_0^{(-1)}(\vec{c})=\bar w_0(\vec{c})$ (proportional to the masses) drive the asymptotics when $\sigma=\pm 1$\footnote{In physical terms, the asymptotic expansion (\ref{Qasy}) can be easily seen as the consequence of that of a 'dressed' relativistic momentum $Z(\theta)\sim \sinh \theta$ satisfying a system of non linear integral equations \cite{FRS}. In this perspective, we have found for $w_0$ an explicit form \cite{FRS}, which simplifies in the case of only one module, $c_0$, into $w_0=-r/[4 \cos \pi/(2N)]$ as consequence of a unique non linear integral equation with form $Z(\theta)= MR \sinh \theta+\dots$ with $r=M R>0$ with $M$ is the (lightest) mass and $R$ the circumference of the cylinder.}. Now, in order to make (\ref {Qasy}) compatible with real-analyticity (\ref {TQreal-an}) and quasi-periodicity, the intuitive condition
\be
\bar w_0 (\bar \vec{c})=w_0(\vec{c}^{R^{-1}})
\label{wper} \, 
\ee
must hold. The next step is to transform the functional equation (\ref {TQstart}) by use of (\ref {qper}, \ref {TQreal-an}, \ref {Qasy}, \ref {wper}) into a handier integral equation
\ba
&& Q_{\pm} \left ( \theta +i \frac{\tau}{2}, \vec{c}\right )= q(\theta, \vec{c})  \pm  \label {Q-masterint} \\
&& \pm \int _{-\infty}^{+\infty} \frac{d\theta '}{4\pi} \tanh \frac{\theta -\theta'}{2} T \left ( \theta ' +i\frac{\tau}{2}, \vec{c} \right )e^{-w_0 (\vec{c}) (e^{\theta} + e^{\theta '} )  -\bar w_0 (\vec{c}) (e^{-\theta} + e^{-\theta '} )}
e^{\pm (\theta -\theta ')l} Q_{\pm} \left ( \theta ' +i\frac{\tau}{2}, \vec{c}\right )
\nonumber \, ,
\ea
valid in the strip $|\textrm{Im} \theta | < \pi$, where the massive field theory driving term
\be
q(\theta, \vec{c})=C_{\pm}
e^{\pm \frac{i\pi}{4}\pm \left ( \theta + \frac{i\pi}{2}\right )l }
e^{-w_0(\vec{c}) e^{\theta}-\bar w_0 (\vec{c}) e^{-\theta}} \, , \quad C_{\pm } \in \mathbb{R} \, ,
\ee
is a consequence of (\ref{Qasy})\footnote{For excited states we expect also $T(\theta)$ to be different with some zeroes.}. Outside the strip $|\textrm{Im} \theta | < \pi$ the functions $Q_{\pm} \left ( \theta +i\frac{\tau}{2}, \vec{c}\right )$ are continued analytically. To prove it, we can 'invert' the $\pm i\pi/2$ shift operator (in the l.h.s. of (\ref {TQstart})) by applying the $\tanh x$ integral kernel by virtue the (residue) relation
\be
\lim _{\epsilon \rightarrow 0^+}\left [ \tanh \left (x+\frac{i\pi}{2}-i\epsilon\right ) - \tanh \left (x-\frac{i\pi}{2}+i\epsilon\right )\right ] =2\pi i \delta (x) \, , \quad x \in \mathbb{R} \, .
\label{shifttan}
\ee
Then, the driving term of (\ref {Q-masterint}) is the zero mode of the shift operator which reproduces the asymptotics (\ref {Qasy}); the shift of half the period guarantees quasi-periodicity (\ref{qper}) provided (\ref{TQreal-an}) and (\ref{wper}) hold.

A crucial fact is now the observation that the integral in equation (\ref {Q-masterint}) converges at large $|\theta|$ only in some circumstances. In fact, we can restrict ourselves to the case $N$ real and only one module, $c_0$, and prove the leading asymptotic expansions by inserting (\ref{Qasy}) respectively inside (\ref {TQstart}) (with $|\textrm{Im}\theta |<\pi/2N-\pi/2$) and (\ref{TQ2}) (with $\pi/2N-\pi/2<\textrm{Im} \theta <0$)
\be
T \left (\theta  + \frac{i\tau}{2} \right ) \simeq - 2 \sin \pi l \, e^{2w_0e^{\theta}+2\bar w_0e^{-\theta}} \, , \ N<1 \ ; \quad  T \left (\theta  + \frac{i\tau}{2} \right ) \simeq  e^{2w_1e^{\theta}+2\bar w_1e^{-\theta}} \, , \ N >1 \, \label{Tlargetheta}\,\, .
\ee
For $N>1$ we had to define $w_1=-\left ( \sin  \frac{\pi }{2N} \right )ie^{\frac{i\pi}{2N}} w_0$, $w_0$ real, from which $\textrm{Re}\, w_1=\left (\sin \frac{\pi}{2N} \right )^2  w_0$ drives the convergence ($\textrm{Re}\, w_1<\textrm{Re}\, w_0$) or the divergence ($\textrm{Re}\, w_1>\textrm{Re}\, w_0$) according to (\ref{Qasy}). Instead, for $N<1$ we witness a striking divergence\footnote{If we assume, on physical grounds, $r=MR\geq 0$ then $w_0=-r/[4 \cos \pi/(2N)]<0$ ({\it cf.} above) for $N>1$, so that $\textrm{Re}\, w_1>0>\textrm{Re}\, w_0$ leads to a divergence, too.}. Finally, the above reasoning applies for sufficiently small moduli $\vec{c}$ as the dependence of $w_0(\vec{c})$ is continuous in $\vec{c}$ \cite{FRS}. However, the form of the integral suggests a regularisation by the introduction of an auxiliary space, which will turn out to be just the independent variables of the ODE side of the correspondence; we will see this in the following.

Let us start from stripping off from $Q_{\pm}$ the model depending terms
\be
C_{\pm} X_{\pm}(\theta, \vec{c})= e^{\mp \frac{i\pi}{4}} e^{\mp \left ( \theta +\frac{i\pi}{2}\right )l} e^{w_0(\vec{c}) e^{\theta}+\bar w_0 (\vec{c}) e^{-\theta}}
Q_{\pm}\left (\theta +i\frac{\tau}{2} , \vec{c}\right )
\label {X-Q conn} \, ,
\ee
so that (\ref {Q-masterint}) becomes the 'universal integral' equation
\be
X_{\pm}(\theta, \vec{c}) =1\pm \int _{-\infty}^{+\infty}\frac{d\theta '}{4\pi} \tanh \frac{\theta -\theta'}{2}
T\left ( \theta ' +i\frac{\tau}{2} ,\vec{c}\right ) E(\theta ',\vec{c})X_{\pm}(\theta', \vec{c}) \, ,
\label{Xint}
\ee
where
\be
E(\theta, \vec{c})=e^{-2w_0(\vec{c}) e^{\theta}-2\bar w_0 (\vec{c}) e^{-\theta}}\simeq
\lim _{\textrm{Re}\,\theta \rightarrow \pm \infty} \frac{Q_{\pm}\left (\theta +i\frac{\tau}{2} , \vec{c}\right )}{Q_{\pm}\left (\theta +i\frac{\tau}{2}-i\pi , \vec{c}\right )} \,
\label{DT}
\ee
depends on the asymptotic behaviour of $Q_{\pm}$. A heuristically crucial point is that it has the form of two plane waves upon identification of $-iw_0(\vec{c})$, $i\bar w_0 (\vec{c})$ with space and of $e^{\pm \theta}$ with momenta variables. In other words, we define the (complex) momenta $\lambda =e^{\theta}$, $\lambda ' =e^{\theta '}$ and make explicit the dependence of $X_{\pm}(\theta, \vec{c})=X_{\pm} (w'_0,\bar w'_0|\lambda)$ on them and on (Wick rotated variables) $w'_0=-iw_0$, $\bar w'_0=i\bar w_0$ to allow its Fourier transform,
\be
K_\pm (w'_0,\xi;\bar w'_0) = \int _{-\infty -i\epsilon}^{+\infty -i\epsilon} d \lambda e^{i(\xi -w'_0)\lambda} [ X_{\pm} (w'_0,\bar w'_0|\lambda)-1] \, , \label {fourtwo2}
\ee
enter the stage. Upon plugging it into equation (\ref {Xint}), we obtain a Volterra equation, valid for $\xi >w'_0$, since (\ref{fourtwo2}) is zero otherwise
\be
K_\pm (w'_0,\xi;\bar w'_0)\pm F(w'_0+\xi;\bar w'_0) \pm \int _{w'_0}^{+\infty} \frac{d\xi'}{2\pi}
K_\pm (w'_0,\xi ';\bar w'_0) F(\xi ' +\xi ;\bar w'_0 )=0 \, .
\label {glm-five2}
\ee
This equation inherits the convergence problems of (\ref{Q-masterint}), but it has almost the structure of a Marchenko-type equation \cite{M}, apart from the form of the 'scattering data'
\be
F(x;\bar w'_0)=i \int _{0}^{+\infty} d\lambda ' e^{-ix\lambda ' +2i\bar w'_0 /\lambda '}
T (\lambda 'e^{i\frac{\tau }{2}}) \, , \label{Fexpr}
\ee
which depend on $w'_0=-iw_0$ (in an intricate way) because $T$ does. Let us now generalise it to the proper form of a Marchenko equation \footnote{Although it is not a crucial issue, we note that for $x,\bar w'_0$ real the integral (\ref {Fexpr}) converges in the case $N>1$ (of interest for applications to Wilson loops/scattering amplitudes). In fact, if we start from the case with only one module, $c_0$, as we have seen above $w_0<0$ and then at large $|\theta|$ the asymptotic behaviour (\ref{Tlargetheta}) is exponentially small thanks to $\textrm{Re}\, w_1=\left (\sin \frac{\pi}{2N} \right )^2  w_0<0$. Then, the convergence still holds if we introduce a set of sufficiently small moduli $\vec{c}$ as the dependence of $w_0(\vec{c})$ is continuous in $\vec{c}$ \cite{FRS}.}. In fact, we can promote the two parameters $w_0(\vec{c})=i w'_0(\vec{c})$, $\bar w_0(\vec{c})=-i \bar w'_0(\vec{c})$ in (\ref {glm-five2}, \ref {Fexpr}) to independent dynamical variables
\be
w=iw', \,\,\,\,  \bar w=-i\bar w' \, ,
\label{ode-im-variables}
\ee
respectively, everywhere except that in the transfer matrix $T$ (which contains $w'_0$ and $\bar w'_0$) which is left unscathed. This amounts indeed to introducing the auxiliary space w.r.t. which the derivatives of the ODE are taken. On the contrary, physical intuition suggests that the dependence of $T$ (generator of the integrals of motion) on $r=MR\propto w_0(\vec{c})$ is to stay unchanged. In other words we are suitably extending the RG parameters $w_0(\vec{c})$, $\bar w_0(\vec{c})$ with a two dimensional space $w',\bar w'$, which eventually will constitute the coordinate space for the ODE side of the correspondence. Besides, the variability of the moduli $\vec{c}$, which control the masses of the model, makes this promotion of $w_0(\vec{c})$, $\bar w_0(\vec{c})$ rather well motivated, although the deep meaning and relevance of the extra space of the ODE/IM in a RG perspective are still to be better investigated. After that, Volterra equation (\ref {glm-five2}) generalises to a Marchenko-like equation \cite{M} for $K_\pm$
\be
K_\pm (w',\xi;\bar w')\pm F(w'+\xi;\bar w') \pm \int _{w'}^{+\infty} \frac{d\xi'}{2\pi}
K_\pm (w',\xi ';\bar w') F(\xi ' +\xi ;\bar w' )=0 \, , \quad \xi >w' \, ,
\label {glm-five}
\ee
where the known term is
\be
F(x;\bar w')=i \int _{0}^{+\infty} d\lambda ' e^{-ix\lambda ' +2i\bar w' /\lambda '}
T (\lambda 'e^{i\frac{\tau }{2}}, \vec{c}) \, . \label{Fexpr2}
\ee
In fact, this is a regularisation of the above mentioned divergences of (\ref{Q-masterint}) as (\ref{glm-five}, \ref{Fexpr2}) can be derived by extending the $Q$-functions $Q(\theta +i\tau/2,\vec{c})$, {\it i.e.} $X_{\pm} (\theta,\vec{c})$, to 'dynamical' counterparts $X_{\pm} (w',\bar w'|\lambda)$ satisfying this well defined extensions of (\ref {Xint})
\be
X_{\pm}(w',\bar w'|\lambda)=1\pm \int _{0}^{+\infty} \frac{d\lambda '}{4\pi \lambda '}
\frac{\lambda -\lambda '}{\lambda + \lambda '}
T (\lambda 'e^{\frac{i\tau}{2}}, \vec{c})e^{-2iw' \lambda ' +2i\bar w ' /\lambda '}
X_{\pm}(w',\bar w'| \lambda ') \label {515ter} \, .
\ee
Then, the Marchenko-like equations (\ref {glm-five}) come out upon Fourier transforming with the definitions
\be
K_\pm (w',\xi;\bar w') = \int _{-\infty -i\epsilon}^{+\infty -i\epsilon} d \lambda e^{i(\xi -w')\lambda} [ X_{\pm} (w',\bar w'|\lambda)-1] \, , \label{fourtwo}
\ee
and it is a regularisation of (\ref{glm-five2}) by {\it moving} the product of mass and length, $r=MR\rightarrow w, \bar{w}$. Thus, it is difficult to immagine this in a theory without mass dependence, but we will obtain CFTs as a scaling limit when $M\rightarrow 0$ in section \ref{cft case}. Obviously, all these manoeuvres have been conceived for real $\xi, w', \bar w'$. Yet, we expect that we can analytical continue in the complex plane to the point $(w',\bar w')=(-iw_0(\vec{c}),i\bar w_0(\vec{c}))$ where the quantities $X_{\pm}$ should give back the integrability functions $Q_{\pm}$. However, this limiting procedure is delicate and to correctly perform it we derive (for complex $w', \bar w'$) Schr\"{o}dinger equations for
\be
\psi _{\pm}(w',\bar w'|\lambda)=X_{\pm} (w',\bar w'|\lambda) e^{-i w' \lambda+i\bar w' \lambda ^{-1}} \,\, ,
\label{psi-X}
\ee
appearing in (\ref {fourtwo}) and with asymptotic plane wave behaviour\footnote{These will result to be just the so-called Jost solutions of the Schr\"{o}dinger equation.}. Actually, the Schr\"odinger equation originates easily from the peculiar form of the Marchenko-like equation (\ref {glm-five}). In fact, we need to use the inverse  Fourier transform of (\ref {fourtwo}) which for $\textrm{Im}\lambda <0$ {\it i.e.} $-\pi<\textrm{Im}\theta <0$ takes the form
\be
X_{\pm} (w',\bar w'|\lambda)-1=\int _{-\infty} ^{+\infty} \frac{d\xi}{2\pi} e^{-i(\xi-w') \lambda} K_\pm (w',\xi;\bar w')=\int _{w'} ^{+\infty} \frac{d\xi}{2\pi} e^{-i(\xi-w') \lambda} K_\pm (w',\xi;\bar w')
\label {fourinv2} \, .
\ee
Then, we differentiate twice (\ref{fourinv2}) and use (\ref {glm-five}) to arrive at the Schr\"odinger equations
\be
\frac{\partial ^2}{\partial {w'}^2} \psi _{\pm}(w',\bar w'|\lambda)+\lambda ^2 \psi _{\pm}(w',\bar w'|\lambda)=u_{\pm}(w';\bar w') \psi _{\pm}(w',\bar w'|\lambda) \, ,
\label {psiSch-}
\ee
with potentials
\be
u_{\pm}(w';\bar w')=-2\frac{d}{dw'} \frac{K_{\pm}(w',w';\bar w')}{2\pi} \, ,
\label {upot}
\ee
determined by the solution of the Marchenko-like equation (\ref {glm-five}). Despite this specific case, the method of passing from a linear integral equation to a differential one holds more in general and thus can be applied to a variety of cases for generalising the correspondence. Now, to cover the domain  \textrm{Im} $\theta >0$ {\it i.e.} $\textrm{Im}\lambda >0$ we need to move the integration straight line in (\ref{fourtwo}) slightly above the real axis and define another Fourier transform
\be
\tilde K_\pm (w',\xi;\bar w') = \int _{-\infty +i\epsilon}^{+\infty +i\epsilon} d \lambda e^{i(\xi -w')\lambda} [ X_{\pm} (w',\bar w'|\lambda)-1] \, \, .
\label{fourtwobis}
\ee
Of course, this leads to Marchenko-like equations with different forms and same known function $F$:
\be
\tilde K_\pm (w',\xi;\bar w')\mp F(w'+\xi;\bar w') \mp \int _{-\infty}^{w'} \frac{d\xi'}{2\pi}
\tilde K_\pm (w',\xi ';\bar w') F(\xi ' +\xi ;\bar w' )=0 \, , \quad \xi <w' \, .
\label {glm-fivebis}
\ee
Thanks to the presence of the same function $F$, the solutions of (\ref{glm-five},\ref {glm-fivebis}) are simply linked by :
\be
K_\pm (w',\xi;\bar w')=-\tilde K_\pm (w',\xi;\bar w') \, \label{KtildeK} \, .
\ee
The inverse of (\ref {fourtwobis}) gives $X_{\pm} (w',\bar w'|\lambda)$ of (\ref {515ter}) for $0<\textrm{Im}$ $\theta <\pi$
\be
X_{\pm}(w',\bar w'|\lambda)-1=\int _{-\infty}^{w'}\frac{d\xi}{2\pi} e^{-i(\xi-w') \lambda }\tilde K_{\pm}(w',\xi;\bar w')
\ee
and the corresponding wave functions (\ref{psi-X}) satisfy, as above, the same Schr\"{o}dinger equation as (\ref {psiSch-}):
\be
\frac{\partial ^2}{\partial {w'}^2} \psi _{\pm}(w',\bar w'|\lambda)+\lambda^2 \psi _{\pm}(w',\bar w'|\lambda)=u_{\pm}(w';\bar w') \psi _{\pm}(w',\bar w'|\lambda) \,, \,\,\, |\textrm{Im} \theta| <\pi \, ,
\label {psiSch}
\ee
with potentials (\ref{upot}). To be precise, for deriving the latter also for {\it real}\, $\theta$ we need to prove a 'dynamical' analogue of the $TQ$ relation, which relates real $\theta$ with imaginary ones (positive and negative). Dubbed $T\psi$ relation, it is obtained by inverting (\ref {515ter}) with (\ref {shifttan})
\be
T\left (\theta +i\frac{\tau}{2}, \vec{c} \right )\psi_{\pm}(w',\bar w'|\theta ) = \mp i \psi_{\pm}(w',\bar w'|\theta +i\pi)\pm i \psi_{\pm}(w',\bar w'|\theta -i\pi) \label{psifunctrel} \, .
\ee
Equations (\ref  {psiSch},\ref {psifunctrel}) can be interpreted with complex $w',\bar w'$ analytically continued  with the aim of establishing the connection between the wave functions and the $Q$-functions. In fact we can derive a partial differential equation for the potentials in (\ref{psiSch}). We introduce a sort of 'complex conjugate' of (\ref{psiSch})
\be
\frac{\partial ^2}{\partial \bar w^{\prime \, 2}} \psi _{\pm}^{bar}(w',\bar w'|\lambda)+\lambda ^{-2}\psi _{\pm}^{bar}(w',\bar w'|\lambda)=\bar u_{\pm}(w',\bar w')  \psi _{\pm}^{bar}(w',\bar w'|\lambda) \,   ,
\label {barpsiSch}
\ee
with $\psi _{\pm}^{bar}(w',\bar w'|\lambda)=\bar \psi _{\pm}(w',\bar w'|\bar \lambda ^{-1})$. This step allows us to reformulate the Schr\"odinger equations (\ref {psiSch}, \ref {barpsiSch}) with potentials
\be
u_{\pm}\bigl (w',\bar w'\bigr)=\pm \frac{\partial ^2}{\partial w^2} \hat \eta  (w,\bar w)- \left ( \frac{\partial }{\partial w} \hat \eta (w,\bar w) \right )^2 \, , \,\,\,\,\, w'=-iw,\,\, \bar w'=i\bar w \label{ueta}
\ee
from the Lax matrix linear problem  (\ref{Laxpr}) and  (\ref {conn}),
whose consistency condition $[D,\bar D]=0$ gives
\be
\frac{\partial ^2}{\partial w \partial {\bar w}} \hat \eta  = 2\sinh 2\hat \eta \, ,
\label{etashg}
\ee
the classical shG equation: this is indeed a very useful tool in asymptotic analysis (for instance around $(w,\bar w)=(w_0(\vec{c}),\bar w_0(\vec{c}))$). Importantly, the potential appearing in the Schr\"odinger problem depends only on the transfer matrix: in fact, it can be obtained by solving (by iterations or Laplace transform) the Marchenko-like equation (\ref{glm-five}, \ref{Fexpr2}) for $K_{\pm}(w',w';\bar w')$ and eventually takes on the following form in terms of Fredholm determinants or tau functions, $\tau _{\pm}$,
\ba
u_{\pm}(w';\bar w')&=&-2\frac{d}{dw'} \frac{K_{\pm}(w',w';\bar w')}{2\pi} =-2\frac{\partial ^2}{\partial { w'} ^2}
\ln \tau _{\pm}(w', \bar w') \label{udiago}\\
\ln \tau _{\pm}(w', \bar w')&=& \ln \det (1\pm \hat V)=\textrm{Tr}\ln (1\pm \hat V) \, , \quad \hat V(\theta , \theta ')=
\frac{T \left (\theta  + i\frac{\tau}{2} , \vec{c} \right )}{4\pi}\frac{ e^{-2iw'e^{\theta }+2i\bar w ' e^{-\theta }}}{\cosh \frac{\theta -\theta '}{2}}
\label{Kdiago}
\ea
Upon carefully inspecting (\ref{ueta}) and (\ref{udiago}), we also obtain an expression for the shG solution
\be
\hat \eta (w,\bar w)= \ln \tau_+ (-iw,i\bar w)-\ln \tau _-(-iw,i\bar w) =
\sum _{n=1}^{+\infty}\frac{2}{2n-1}
\int \prod _{i=1}^{2n-1} \frac{d\theta_i}{4\pi} T \left (\theta _i + \frac{i\tau}{2} , \vec{c} \right ) \frac{ e^{-2we^{\theta _i}-2\bar w e^{-\theta _i}}}{\cosh \frac{\theta _i -\theta _{i+1}}{2}} \, .
\label{etatau}
\ee
Thus we find a confirmation of correctness of our inverse derivation as the solution (\ref {etatau}) was already presented in \cite {LZ}. Nevertheless, the domain of validity of the equation (\ref{glm-five}, \ref{Fexpr2}) and its solution (\ref {etatau}) depends on the behaviour of $T$ at large $\textrm{Re}\, \theta$. In particular, (\ref{Tlargetheta}) allow us to frame the situation at least for $N$ real and small moduli except $c_0$: if $N<1$ the multi-integrals converge for any $\textrm{Re} \, w>\textrm{Re} \, w_0$, whereas for $N>1$ they converge only for $\textrm{Re}\, w>\textrm{Re} \, w_1> \textrm{Re} \, w_0$, the last inequality holding because -- as proven above -- $\textrm{Re}\, w_1=\left (\sin \frac{\pi}{2N} \right )^2 w_0$ and $-r \sim w_0<0$. In other words, when $N>1$ the convergence stops at $w_1$ before $w=w_0$ and the equations need to be analytically continued. Yet, the shG equation (\ref{etashg}) remains valid. In fact, we can derive a relevant limit value by using the results for $N<1$ and $\textrm{Re} \, w>\textrm{Re} \, w_0$: when $w\to w_0$ we can substitute $T$ in (\ref {etatau}) with its leading expansion, first of (\ref {Tlargetheta}), which contains $l$. Then, we absorb the phase of $w-w_0$ by a shift in $\theta _i$ and follow a general method described in \cite{SMI}: in the $n$-th term we change integration variables in $\alpha_i=\theta_{i+1}-\theta_1, i=1,\dots, 2n-1$, integrate on $\theta_1$ to obtain the Bessel $K_0$, expand it (at leading order) for small argument proportional to $|w-w_0|\rightarrow 0$, integrate on all the remaining $\alpha_i$ and eventually resum the series into
\be
\hat \eta (w,\bar w) \underset{w \to w_0}{=} 2l \ln |w-w_0| +... \, . \label{smallw}
\ee
Importantly this formula does not contain $N$. Of course, it satisfies the shG equation  (\ref{etashg}), independent of $N$, too: hence, it is very natural to assume it holds for generic $N$ (and small moduli $\vec{c}$). It coincides with the asymptotic behaviour assumed in \cite{LZ} for the derivation of the functional and integral equations\footnote{Similarly, if $N>1$  we can approximate $T$ in (\ref{etatau}) with the second of (\ref {Tlargetheta}) to obtain the limit $w\rightarrow w_1$ ($\textrm{Re} \, w_1> \textrm{Re} \, w_0$): $\hat \eta (w)= -\frac{1}{3}\ln |w-w_1|+\dots$\,.}.

As anticipated, for the opposite direction of the correspondence to hold, we need to prove a formula which recovers the $Q$-functions from the wave functions (supposedly, in the form of a limiting value $w\rightarrow w_0$). Explicitly, we are going to prove that this is realised by this 'regularisation'
\be
 \lim _{w\rightarrow w_0} (w-w_0)^{\pm l}\psi _{\pm}(w',\bar w'|\theta)= C e^{\mp \theta l}Q_{\pm} \left (\theta + i \frac{\tau}{2}, \vec{c} \right )  \, ,\,\, w'=-iw, \,\, \bar w'=i\bar w  ,
 \label{psiQrel}
\ee
with a constant $C$ to be chosen. In fact, at the leading order (\ref{smallw}) the potentials (\ref{ueta}) simplify to $u_{\pm} \simeq - l(l\pm 1)/(w-w_0)^2$. This entails that equation $(\ref{psiSch})$ with $+$ has two solutions labelled by $\alpha =-l, l+1$ behaving as $f_+^{(\alpha)}(w',\bar w') \sim (w-w_0)^{\alpha}$ when $w\rightarrow w_0$, whilst equation with $-$ has other two solutions labelled by $\alpha =l, -l+1$ behaving as $f_-^{(\alpha)}(w',\bar w')\sim (w-w_0)^{\alpha}$. Of course, $\psi _{\pm}(w',\bar w'|\lambda)$ can be expressed linearly in terms of the $f_\pm^{(\alpha)}(w',\bar w')$ as
\ba
\psi _+(w',\bar w'|\lambda)&=&B_+(\theta)  f_+^{(l+1)}(w',\bar w')+
A_+(\theta) f_+^{(-l)} (w',\bar w') \nonumber \\
\psi _-(w',\bar w'|\lambda)&=&A_-(\theta) f_-^{(l)}(w',\bar w')+
B_-(\theta) f_-^{(-l+1)}(w',\bar w') \, . \label{connex}
\ea
Since $-1/2<l<1/2$, in the limit $w\rightarrow w_0$ the first $(+)$ is dominated by $f_+^{(-l)}\sim (w-w_0)^{-l}$ while the second $(-)$ by $f_-^{(l)}\sim (w-w_0)^l$. Then, the relations (\ref {psiQrel}) pick up in both cases the finite quantities $A _{\pm}(\theta)$. Upon identifying the latter respectively with $e^{\mp \theta l}Q_{\pm} (\theta + i \tau /2, \vec{c})$ up to a multiplicative constant, these $Q_{\pm} (\theta + i \tau /2, \vec{c})$ satisfy the relevant relations: 1) the {\it universal} $TQ$-system (\ref {TQstart}) thanks to the $T\psi$ relation (\ref {psifunctrel}); 2) the expansion (\ref {Qasy}) as given by large $\theta$ analysis on the Schr\"odinger equation (\ref{psiSch})\footnote{This can be achieved either via a careful large energy expansion with attention to the divergence for $w\rightarrow w_0$ as in \cite{FG1} or via the series solution of the Fredholm integral equation below (\ref{Xtheta}) for the wave function (\ref{psi-X}).}. In conclusion, (\ref {psiQrel}) is proven and connects the solutions of the $QQ$ relation (\ref{Qw}) to the solutions of the Schr\"odinger equations (\ref{psiSch}), constructed from the former and circumventing the divergence problem of equations (\ref {Q-masterint}, \ref {Xint}).

As an interesting consequence, our method expresses the potentials of the Schr\"odinger equations (\ref{psiSch}) -- solutions of integrable partial differential equations -- in terms of Fredholm determinants, via a Marchenko-like equation, and then in terms of functional equations. Actually, more in general, it links Fredholm's to Sturm-Liouville's theory as it gives explicit and useful forms for solutions to the latter in terms of integral equations of the former: upon finding the solutions to the Marchenko-like equations (\ref {glm-five},\ref {glm-fivebis}) (as done for the diagonal part, the tau-functions (\ref{Kdiago})), we invert (\ref{fourtwo}, \ref{fourtwobis}) and combine the two in (\ref {psifunctrel}) to obtain an integral equation for the wave-function (\ref{psi-X})
\be
E(\theta )X_{\pm}(\theta )=  -2 E(\theta )\mp \int _{-\infty}^{+\infty}\frac {d\theta '}{4\pi} \frac{e^{-v(\theta)-v(\theta ')}}{\cosh \frac{\theta -\theta'}{2}}E(\theta ') X_{\pm}(\theta ')\, , \, \,\,\, |\textrm{Im} \theta| <\pi \, ,
\label{Xtheta}
\ee
with the definitions $\lambda=e^{\theta}$ and
\be
E(\theta)=\sqrt{2}e^{-v(\theta)}e^{\frac{\theta}{2}} \, , \quad e^{-2 v(\theta)}=e^{-2iw'e^{\theta}+2i\bar w' e^{-\theta}}
T\left ( \theta + i\frac{\tau}{2}, \vec{c}\right ) \, .
\ee
It allows for a solution as recursive series in these functions and $1/\cosh$ kernel, which is very much complementary to ODE/IM treatment and results \cite{FRnext}.



\section{The conformal limit}
\setcounter{equation}{0}
\label{cft case}

In the present construction the field $\hat \eta$ is invariant under $\Omega$-symmetry (recall and compare with (\ref{crot})): $w\rightarrow -w e^{\frac{i\pi}{N}}, \vec{c}\rightarrow \vec{c}^R, \theta \rightarrow \theta -\frac{i\pi}{N}$. Indeed,  $\hat \eta (-we^{\frac{i\pi}{N}},-\bar w e^{-\frac{i\pi}{N}},\vec{c}^{R})=\hat \eta (w,\bar w,\vec{c})$ derives  from (\ref{etatau}) thanks to the periodicity of $T$ (\ref {Tper}). Of course, this invariance extends to the potentials (\ref{ueta}, \ref{psiSch}) ($w=iw'$, $\bar w=-i\bar w'$) which maintain the same form with 'rotated' $\psi_\pm$ (broken symmetry). This simple symmetry means that $\hat \eta$ may be re-interpreted as depending on a {\it natural} variable $z$ defined, for instance, by means of the polynomial $p(z,\vec{c})=z^{2N}+\sum_{n=0}^{2N-2}c_nz^n$ and the differential
\be
\frac {d\bar w}{d\bar z}=\sqrt {\bar p(z)} \, , \quad \frac {dw}{dz}=\sqrt {p(z)} \, , \quad w(z=0)=w_0 \,   \, .
\label {chng}
\ee
In fact, to make manifest the $\Omega$-symmetry in the new variable $z\rightarrow ze^{\frac{i\pi}{N}}$, $\vec{c}\rightarrow \vec{c}^R$, the function $p$ may be left invariant: $p(z,\vec{c})=p(ze^{\frac{i\pi}{N}},\vec{c}^R)$, and a polynomial is the simplest choice. Let us focus on the $+$ case of (\ref {psiSch}) as the $-$ sign goes {\it pari passu}. In the new variable we obtain a gauge equivalent wave function $\psi_{+} (w')= {(p(z))}^{1/4}  \tilde \psi _{+}(z)$ satisfying the {\it modified} Schr\"odinger equation
\be
-\frac {d^2}{dz^2} \tilde \psi _{+}+e^{2\theta}p(z) \tilde \psi _{+}-p(z) \left ( \frac{1}{{(p(z))}^{1/4}}  \frac{d^2}{d w^2} (p(z))^{1/4}+u_{+} \right )  \tilde \psi _{+}=0 \,  \label{zdiffeq}\,\, ,
\ee
whose potential is given by (\ref{ueta}) in the simple form (not equally elegant for the $-$ case) 
\be
p(z) \left ( \frac{1}{{(p(z))}^{1/4}}  \frac{d^2}{d w^2} (p(z))^{1/4}+u_{+} \right )= \frac{\partial ^2}{\partial z^2} \eta  (z,\bar z)- \left ( \frac{\partial }{\partial z} \eta (z,\bar z) \right )^2  \, ,
\ee
in terms of $\eta =\hat \eta +\frac{1}{16} \ln p \bar p$, solution to the {\it modified} shG equation
\be
\partial _z \partial _{\bar z} \eta =e^{2\eta}-p(z)\bar p(z) e^{-2\eta} \, . \label{modshg}
\ee
At this stage we can deduce  the $z, \bar z\rightarrow 0$ asymptotics, when $p(z)\simeq c_0$ entails small deviation  $w-w_0\simeq \sqrt {c_0} \, z$ and then, from (\ref{smallw}), the elegant outcomes 
\be
\eta = \hat \eta +\frac{1}{16} \ln p \bar p = l \ln (z\bar z) +O(1) \Rightarrow p(z) \left ( \frac{1}{{(p(z))}^{1/4}}  \frac{d^2}{d w^2} (p(z))^{1/4}+u_{+} \right )= - \frac{l(l+1)}{z^2}+O(1) \, .
\ee
In fact, last formula is enough to deduce the 2D CFT case \cite{ODE-IM} as given by the static case $\bar z=0$ along with the scaling limit $\theta \rightarrow +\infty$ with $x,\tilde c_n$ fixed and driving as $z=xe^{-\theta/(1+N)}$ and $c_n=\tilde c_ne^{-\theta (2N-n)/(1+N)}$ go to zero\footnote{Hence, the asymptotic (\ref{smallw}) of shG equation is the reason of simplicity of the original conformal ODE/IM. Yet, this regime is static ($\bar z =0$) and thus miss the general scenario of Lax pairs and shG evolution.}. Eventually, $\tilde \psi _+(z)$ goes into the CFT limit $\psi ^{cft}_+(x)$ and the Schr\"{o}dinger equation (\ref {zdiffeq}) acquires polynomial potentials
\be
-\frac {d^2}{dx^2} \psi ^{cft}_+(x)+\left (p(x,\vec{\tilde c})+\frac {l(l+1)}{x^2} \right) \psi ^{cft}_+(x)=0 \, , \quad
p(x,\vec{\tilde c})=x^{2N}+\sum_{n=0}^{2N-2}\tilde c_n x^n \, ,
\label{cftsch}
\ee
This reproduces and extends known results of the conformal case \cite {DDTSUZ} and, in particular, with only one modulus $\tilde c_0=-E$, $\tilde c_n=0, n\geq 1$, it reduces to that in \cite{ODE-IM} \footnote{Then the $Q$s become entire functions of the energy $E$ because of quasi-periodicity (\ref{qper}) and analyticity at $E=0$.}.

\section{A relevant example}
\setcounter{equation}{0}
\label{ex}

In the model $l=0$, $N=1/2$ $ (\tau =3\pi)$ with only the real module $c_0$ ({\it i.e.} $w_0=\bar w_0$) calculations can be made even more explicit. In fact, the $QQ$-system and the quasi-periodicity read
 \be
 Q_+(\theta )Q_-(\theta +i\pi ) + Q_-(\theta  )Q_+(\theta +i\pi )=-2 \, , \quad
 Q_{\pm}(\theta +3i\pi)=\pm i Q_{\pm}(\theta)
 \ee
and, as a consequence, the transfer matrix
\be
T(\theta )=\frac{i}{2} \left [ Q_+(\theta +i\pi )Q_-(\theta -i\pi )- Q_+(\theta -i\pi )Q_-(\theta +i\pi ) \right]=1 \, .
\ee
When $T$ is a constant, the tau functions $\tau _{\pm}$ (\ref {Kdiago}) enjoy {\it radial symmetry}, namely they depend only on $t=4\sqrt{w'\bar w'}$ (and not on the phase $\phi$ of $w=\frac{t}{4} e^{i \phi}$) and read
\be
\ln \tau _{\pm}(t)=\sum _{n=1}^{+\infty} \frac{(-1)^{n+1}(\pm 1)^n}{n}
\int  \prod _{i=1}^n \frac{d\theta_i}{4\pi } \frac{ e^{-t\cosh \theta _i}}{\cosh \frac{\theta _i -\theta _{i+1}}{2}} \, .
\label {tauti}
\ee
Analogous dependence holds for the solution $\hat \eta $ (\ref {etatau}) of the shG equation
\be
\hat \eta (t)=\sum _{n=1}^{+\infty}\frac{2}{2n-1}
\int \prod _{i=1}^{2n-1} \frac{d\theta_i}{4\pi} \frac{ e^{-t \cosh \theta _i}}{\cosh \frac{\theta _i -\theta _{i+1}}{2}} \, , \label{etataut0}
\ee
which then satisfies the Painlev\'{e} III$_3$ equation\footnote{This a particular solution depending at most on one parameter, the value of $T$.}
\be
\frac{1}{t} \frac{d}{dt} \left ( t \frac{d}{dt} \hat \eta (t) \right ) = \frac{1}{2} \sinh 2\hat \eta (t) \, .
\label{pain3}
\ee
It is an instructive exercise to compute (\ref {tauti}) at small $t$ by the method used to derive (\ref{smallw}), {\it i.e.} upon shifting  $\alpha_i=\theta_{i+1}-\theta_1, i=1,\dots, n$ \cite{SMI}: $\ln \tau _{\pm}(t)\simeq d_{\pm} \ln t$, $d_{\pm}=(1\mp 6)/36$. Accordingly, the potentials (\ref {udiago}) 
for small $w'$ expand as
\be
u_{\pm}(w')=-2 \frac {\partial ^2}{\partial w'^2}\ln \tau _{\pm}(t)=\frac{d_{\pm}}{w'^2}+ ...  \, ,
\label{u+limit}
\ee 
whose leading term is just the CFT limit. In fact, the change of variables (\ref{chng}) takes the explicit form
$w=\frac {2}{3}(z+c_0)^{\frac {3}{2}}$, $\bar w=\frac {2}{3}(\bar z+c_0)^{\frac {3}{2}}$ and the limit amounts to consider 
\be
\bar w'=\frac{2}{3}E^{3/2}e^{-\theta} \, , \quad w'=\frac{2}{3}(-x+E)^{3/2}e^{-\theta} \, , \quad  \psi _{\pm}(w',\bar w'|\lambda)=\left(w'e^{\theta}\right)^{\pm 1/6} \psi ^{cft}_{\pm}(x) \, , \quad \theta \rightarrow +\infty \, ,
\label{cftlim}
\ee
with $x,E=-\tilde c_0$ fixed, namely $t=4\sqrt{w'\bar w'}\rightarrow 0$. Very inspiring are the conformal versions of (\ref {psiSch}) which take the {\it bispectral} form
\be
-\frac{d^2}{dx^2} \psi ^{cft}_+(x) +(x-E) \psi ^{cft}_+(x)=0 \, ,
\label {airyplus}
\ee
($l=0,N=1/2$ case of (\ref {cftsch})) solved by the Airy function $Ai(x-E)$, and
\be
-\frac{d^2}{dx^2} \psi ^{cft}_-(x) + (x-E) \psi ^{cft}_-(x) + \frac{1}{x-E}\frac{d}{dx}\psi ^{cft}_-(x)=0 \, ,
\label {airyminus}
\ee
solved by its derivative $Ai'(x-E)$ \cite{ODE-IM}. In fact, these can be simply converted into ODE in the spectral parameter $E$\footnote{Here the situation is even simpler as $x$ and $E$ can swap places.}, and hence for $\psi _{\pm}^{cft}(x=0)=Q_{\pm}$ (without divergence).

\subsection{Bispectrality}

Furthermore, for $T$ constant the wave-functions $\psi _{\pm}(w',\bar w'|\lambda)$ actually depend only on two variables $w'\lambda$ and $ \bar w' \lambda ^{-1}$ or $t$ and $\theta +i\phi$ in polar coordinates ($\lambda=e^\theta$). This means that we can swap  the differential equations in $w'$ and $\bar w'$ (\ref {psiSch},\ref{barpsiSch}) with two in $s=e^{\theta+i\phi}$ and $t$
\be
s ^2 \frac {\partial ^2 \psi _{\pm}}{\partial s ^2}+ \left (1-F_{\pm}(t,s) \right ) s \frac {\partial \psi _{\pm}}{\partial s }\mp \frac {t}{2}\frac {d\hat \eta}{dt}F_{\pm}(t,s ) \psi _{\pm}
+\frac {t^2}{16}\left (-s ^2-s ^{-2}+2\cosh 2\hat \eta -4 \left (\frac {d\hat \eta}{dt} \right )^2 \right )\psi _{\pm}=0 ,
\label{bispeq}
\ee
\be
t^2 \frac {\partial ^2 }{\partial t^2}\left (e^{\pm \frac {\hat \eta}{2}}\psi _{\pm}\right )\mp t^2 \frac {d \hat \eta}{dt} \frac {1}{F_{\pm}(t,s )}
\frac {\partial }{\partial t}\left ( e^{\pm \frac {\hat \eta}{2}}\psi _{\pm}\right )+ \frac {t^2}{16}\left [-s ^2 -s ^{-2}-2\cosh 2\hat \eta \right ] e^{\pm \frac {\hat \eta}{2}}\psi _{\pm} =0 \, , 
\label {geneqt}
\ee
with coefficients $F_{\pm}(t,s )=(s e^{\pm \hat \eta}+s ^{-1}e^{\mp \hat \eta})(s e^{\pm \hat \eta}-s ^{-1}e^{\mp \hat \eta})^{-1}$. In other words the problem becomes {\it bispectral} (in differential form) and (\ref {bispeq}) yield the off-critical versions (in the conformal limit: $s t=\frac {8}{3}(E-x)^{\frac {3}{2}}$ with $s\rightarrow +\infty$ and $t\rightarrow 0$) of the (bispectral) ODEs in $E$ stemming from (\ref {airyplus}, \ref {airyminus}) by swapping $x\leftrightarrow - E$. Moreover, we can compute (\ref{bispeq}) in $(w,\bar w)=(w_0,\bar w_0)$ for $\phi =0$ upon shifting $\theta \rightarrow \theta-3i\pi/2$, so that they become ODEs for $\psi _{\pm}(w_0,\bar w_0|\lambda)=Q_{\pm}(\theta)$ (simply (\ref{bispeq}) with the substitution $t\rightarrow r$). An ODE for the $Q$ is a rather desirable situation for quantum integrable systems and in this case the coefficients are in terms of a Painlev\'e transcendent; a similar method will be exploited in the next section on spin chains.

Furthermore, the solutions  $\tau_\pm$ and $\hat \eta$ enjoy radial symmetry, but in the displaced variable $w-w_0=\frac {t}{4} e^{i \phi}$ ($\phi$ independence), even if $N\rightarrow +\infty$ with only one non-zero module $c_0$ (at fixed $l$ and $w_0=-r/4$)\footnote{In fact, the $\Omega$-symmetry sector reduces to a straight half-line.} and anew the problem becomes bispectral. As above the wave functions depend only on two polar coordinates $t$ and $\theta +i\phi$ and equations (\ref {bispeq}, \ref {geneqt}) still hold\footnote{They are linear Lax associated problems for the Painlev\'e $III_3$ equation.}, albeit now with a different definition of the time $t$ and a general  solution $\hat \eta(t)$ of Painlev\'{e} III$_3$ equation (\ref {pain3}) depending on two parameters, $l$ and $r$. At the (movable) pole $t=r$ ($w=0=w_1$), the function $\hat \eta(t)$ expands as
\be 
\hat \eta (t) = -\ln |t-r| +\ln 2- \frac {t-r}{2r}+ \frac {7-16u}{24 r^2}(t-r)^2 + O(t-r)^3 \label
{etalargealfa2}
\ee
with $u$ a constant depending on $r,l$. As a consequence, we obtain for the $Q$-functions, given by $Q_{\pm}\left (\theta + \frac {i\pi}{2} \right )= \lim \limits _{\substack {t\rightarrow r \\ \phi \rightarrow \pi}} [s e^{\pm \hat \eta}-s ^{-1}e^{\mp \hat \eta}]^{-\frac{1}{2}} \psi _{\pm} \left (\theta \right )$ in this limit case, the modified Mathieu equation
\be 
\frac{\partial ^2  Q _{\pm}(\theta)}{\partial \theta ^2} +\left (\frac {r^2}{8}\cosh 2\theta - u \right ) Q _{\pm}(\theta)=0\,.
\ee
It has isomonodromic deformations given by the Painlev\'e $III_3$ equation above (with solution fixed by the parameters $(l, r)$ or $(r,u)$) and monodromies studied in \cite{FG1} in connexion with supersymmetric gauge theories. It also coincides with the $N\rightarrow +\infty$ limit of the $TQ$ (\ref{TQ2}) \cite {LUK}, as can be seen upon improving the second equality (\ref{Tlargetheta}) by sub-leading term $e^{2w_1e^{-\theta}+2\bar w_1e^{\theta}}$ \cite{FRnext}. So that the present development may pave the way for the paramount open problem of studying the monodromies of a finite difference equation, like the $TQ$-relation, and somehow identify the shG equation (\ref{etashg}) as a generalisation of the Painlev\'e isomonodromies. In \cite{FG2} a seed has been sown by finding the connexion between the Floquet and the ODE/IM bases of eigenfunctions.

\section{Spin chains: a particular point of the XXZ}
\label{spin}
\setcounter{equation}{0}

The $TQ$-system for the XXZ spin chain with anisotropy $\Delta=\cos 2\eta$ may be written as $T^{XXZ}(u) Q^{XXZ}_{\pm}(u) =\left [ \sinh (u-i\eta)\right ]^L Q^{XXZ}_{\pm}(u+2i\eta)+\left [ \sinh (u+i\eta)\right ]^L Q^{XXZ}_{\pm}(u-2i\eta)$ and coincides, apart from the extra-factors $[\sinh (u\mp i \eta) ]^L$, with the universal $TQ$-system (\ref {TQstart}) provided $\theta =i\pi u/2\eta$. These extra-factors need some technical adaptation of our procedure, but the general idea is to reproduce the $TQ$-system with the $T\psi$-system (\ref {psifunctrel}) at particular values of the extra variables $w'$ and $\bar w'$. If presented in general, this would require too much space. Yet, an idea of what happens can be gained by considering $\eta =\pi/3$. In fact at this point, for odd number of sites $L=2n+1$, these modifications of $Q$s, $f_\pm(u)=(\sin u)^{2n+1}Q^{XXZ}_{\pm}(u)$, satisfy the simple functional relations $f_\pm (u+2\pi/3)+f_\pm(u-2\pi/3)+f_\pm(u)=0$ \cite{STR}.
In fact, these relations can be reproduced, by the $T\psi$-system (\ref {psifunctrel}) for $N=1/2$, $l=0$, once we define $\tilde \psi _{\pm} =e^{\pm \frac{3iu}{4}} \psi _{\pm}$:
\be
\tilde \psi_{\pm}(w',\bar w'|u+\frac{2\pi}{3})+\tilde \psi_{\pm}(w',\bar w'|u-\frac{2\pi}{3})+\tilde \psi _{\pm}(w',\bar w'|u)=0
\label{newfrel} \, .
\ee
In the end, $\tilde \psi_\pm$ are to be considered at particular values of $w'$ and $\bar w'$ such that the large $u$ behaviour coincides with the one of $f_\pm \sim e^{\frac {3L\mp 1}{2}iu}$ respectively. This is the limit $t \rightarrow 0$ of polar coordinates $w'=-\frac {it}{4} e^{i\phi}$, $\bar w'=\frac {it}{4}e^{-i\phi}$, when $\tilde \psi_\pm$ lose their 'double exponential' behaviour $\sim \exp (iw'\lambda -i\bar w' \lambda ^{-1})$. In detail, we need to look in general for solutions with specific leading behaviour:
\be
\tilde\psi _{\pm}\left (t,\phi |u \right)= t^{L} f_{\pm}\left( u+\frac{2}{3}\phi\right)+...\,\, ,
\label{psifacto}
\ee
where the dependence only on $3u+2\phi$ is a simplification of this case $\eta =\pi/3$. It is peculiar of this case the further possibility of simplifying  (\ref {psiSch}) as $t \rightarrow 0$ with the help of (\ref{u+limit})
\be
 e^{-3iu-2i\phi} \left ( \frac{t^2}{4} \frac{\partial ^2 \tilde\psi _{\pm}}{\partial t ^2} -\frac{t}{4} \frac{\partial  \tilde\psi _{\pm}}{\partial t}-
\frac{1}{4} \frac{\partial ^2 \tilde\psi _{\pm}}{\partial \phi^2}+\frac{i}{2} \frac{\partial  \tilde\psi _{\pm}}{\partial \phi}
-\frac{it}{2} \frac{\partial ^2 \tilde\psi _{\pm}}{\partial t \partial \phi}-d_{\pm} \tilde\psi _{\pm} \right )= \frac {t^2}{16}\tilde\psi  _{\pm} \,\, 
\label{dieq}
\ee
Then, for real $u$ and Bethe roots, hence $f_{\pm}$, we subtract to (\ref {dieq}) its complex conjugate
\be
-2 \sin (3u+2\phi) \left [ \left (\frac{L^2}{4}-\frac{L}{2}-d_{\pm} \right  ) f_{\pm}-\frac{1}{4} \frac{\partial ^2 f_{\pm}}{\partial ^2 \phi} \right ] +2\cos (3u+2\phi) \frac{(1-L)}{2}\frac{\partial f_{\pm}}{\partial \phi} =0  \label {diffeqfpm}
\ee
and obtain a differential equation which does not contain $t$ any more. Eventually, thanks to the particular dependence on $3u+2\phi$, derivation on $\phi$ can be traded for derivation on $u$:
\be
\frac{d^2f_\pm}{du^2}-6n \cot (3u+2\phi)\frac{df_\pm}{du}+ (c_\pm-9n^2)f_\pm=0 \, , 
\label{eqstr}
\ee
with $c_\pm=5/2\mp3/2$ and $L=2n+1$. These equations are the same as (13) of \cite{STR} with $n$ integer yielding odd number of sites $L=2n+1$ ($f_\pm$ are there called $f$ and $g$). In conclusion, for $\phi=0$ they are solved by the forms $f_{\pm}(u)=(\sin u )^{2n+1} P_{\pm}(\cos u)$, with $P_\pm$ polynomials of degree $n$ ($+$ case) or $n+1$ ($-$ case). On the other hand, we know from above that these polynomials must be the $Q$-functions $Q^{XXZ}_{\pm}(u)=P_{\pm}(\cos u)$ for the ground state of the XXZ spin chain with anisotropy $\Delta =-1/2$ ($2n+1$ sites). Arguably this procedure can be extended to generic cases with some dedicated labour \cite{FRnext}.

\section{Final considerations and perspectives.}

We have presented a general procedure to find the solutions $Q_\pm$ of the $QQ$ relation (\ref{Qw}) or $TQ$ relation (\ref{TQ2}) with suitable analytic and asymptotic properties, as from the limit (\ref {psiQrel}) of the respective wave-functions (\ref{psiSch}) around the 'origin' $w_0$. In brief details, we have converted the quantum universal $TQ$ relation (\ref{TQstart}) into the Marchenko-like equation (\ref {glm-five}), written in terms of $T$ only. The latter involves the independent variables $w$ and $\bar w$ (\ref{ode-im-variables}) of the ODE/IM correspondence, crucially introduced as extension of the adimensional renormalisation group parameter $r=MR$. In fact, they are those in which the Schr\"odinger (\ref{psiSch}) can be derived from the Marchenko-like equation. Analogously, also another Schr\"odinger equation (\ref {barpsiSch}) in $\bar w$ holds, giving the compatibility of the two the crucial constraint on the 'potential' (\ref{etashg}). In a nutshell, we have mapped a quantum integrable model into a classical one. 

We have elaborated the procedure in the quantum homogenous sG model (sG with many masses) and found the expected classical shG equation, agreement in the CFT limit (Section \ref {cft case}) and in two peculiar cases leading to the radial symmetric shG, {\it i.e.} Painlev\'e $III_3$ equation (transfer matrix $T=1$ and $N=\infty$, Section \ref {ex}). But our method is so general that, it is supposed to be applicable to many quantum integrable models; the application to the XXZ chain (Section \ref {spin}) goes in that direction by changing the form of the $QQ$- and $TQ$-systems with the same number of $Q$s. Similarly, we can immagine to start from more $QQ$ relations with form analogous to (\ref{Qw}) (higher rank algebras than $su(2)$). As a byproduct, it can be applied to the ordinary scattering relations as a new way to derive the usual Marchenko equation \cite{M}.

In the end we can obtain a variety of potentials with moduli of physical meaning \cite{DDTSUZ, GMN, Ys, DDNT, FRS}  and give an interpretation to the ODE/IM variables (\ref{ode-im-variables}): this make us think that our construction is a powerful tool for charting the space of $2d$ (integrable) theories. In fact, we are able to have control of the CFT regime, i.e. the ultraviolet limit of the RG flow. This spanning may reveal itself also more interesting in view of correspondences of 2d theories with higher dimensional theories, like, for instance, between the $Q$-, $T$- functions and the periods of $\mathcal{N}=2$ SYM \cite{FG1, FG2, IS-17, FGS}.

\vspace{1truecm}

{\bf Acknowledgements}
We thank R. Tateo for important comments and D. Gregori, S. Lukyanov, H. Shu and A. Zamolodchikov for discussions. This work has been partially supported by the grants: GAST (INFN), the MPNS-COST Action MP1210, the EC Network Gatis and the MIUR-PRIN contract 2017CC72MK\textunderscore 003.

\end{document}